\documentclass{INTERSPEECH2023}

\interspeechcameraready

\title{Improving Frame-level Classifier for Word Timings with Non-peaky CTC in End-to-End Automatic Speech Recognition}
\name{Xianzhao Chen, Yist Y. Lin, Kang Wang, Yi He, Zejun Ma}
\address{ByteDance}
\email{\{chenxianzhao, yist.lin0, wangkang, heyi.hy, mazejun\}@bytedance.com}

\begin{document}

\maketitle

\begin{abstract}
End-to-end (E2E) systems have shown comparable performance to hybrid systems for automatic speech recognition (ASR).
Word timings, as a by-product of ASR, are essential in many applications, especially for subtitling and computer-aided pronunciation training.
In this paper, we improve the frame-level classifier for word timings in E2E system by introducing label priors in connectionist temporal classification (CTC) loss, which is adopted from prior works, and combining low-level Mel-scale filter banks with high-level ASR encoder output as input feature.
On the internal Chinese corpus, the proposed method achieves 95.68\%/94.18\% compared to the hybrid system 93.0\%/90.22\% on the word timing accuracy metrics.
It also surpass a previous E2E approach with an absolute increase of 4.80\%/8.02\% on the metrics on 7 languages.
In addition, we further improve word timing accuracy by delaying CTC peaks with frame-wise knowledge distillation, though only experimenting on LibriSpeech.
\end{abstract}
\noindent\textbf{Index Terms}: automatic speech recognition, connectionist temporal classification, word timing, end-to-end

\section{Introduction}

Word timing is the start and end time of a word.
It is widely used in many ASR applications, such as computer-aided pronunciation training, subtitling and filler word removal.
Word timings can also be converted into frame-level classifications for pre-training acoustic model in ASR systems~\cite{huang21e_interspeech}.
Correspondingly, the word timing metric can be used to measure the performance of the frame-level classifier~\cite{sainath2020emitting}.

Recently, E2E systems adopting Deep Neural Networks (DNN) have demonstrated excellent ASR performance~\cite{karita2019comparative,gulati2020conformer}.
Unlike traditional hybrid systems, E2E systems directly align inputs and outputs without modeling phonetic alignment.
In E2E systems, Listen, Attend and Spell (LAS)~\cite{chan2016listen} uses attention mechanisms, CTC~\cite{graves2006connectionist} and RNN Transducer (RNN-T)~\cite{graves2012sequence} introduce blank token to align variable-length inputs and outputs.
Though E2E systems trained with soft alignments achieve superior ASR performance compared to traditional hybrid systems, the estimated word timing in such systems is usually not comparable with that in traditional ones.

There are several methods for estimating word timings in hybrid systems and E2E systems.
In a hybrid system, such as DNN-Hidden Markov Models (HMM) system, a DNN model is trained with frame-level classification using alignments from a pretrained GMM-HMM system, and the HMMs are trained using the posterior probabilities of acoustic features provided by DNN~\cite{kaldi}.
Accurate word timings are obtained from forced alignment of the recognition results. 
However, a hybrid system requires many steps of training, making it less desirable.

In E2E systems, word timings can be estimated by the forced alignment results of character-level CTC models, where the CTC peak of the first character indicate the word start time and the CTC peak of the last character indicate the word end time~\cite{baevski2020wav2vec}. 
The CTC model cannot estimate word timings well when the duration of the modeling unit is relatively long, e.g., Chinese characters.
Because the blank probability of CTC model is dominant in almost all frames, and the non-blank probability is only relatively high in few frames. This is called the peaky behavior~\cite{tian2022peak}.
CTC-based alignments for word timings can be improved by alleviating the peaky behavior~\cite{teytaut2022study, liu2018connectionist}, but these methods have complicated regularization terms which require modifications of the model architecture or CTC loss topology respectively.
A two-pass model trained with frame-level classification, where alignments are from traditional hybrid systems, predicts word timings by constrained attention weights~\cite{sainath2020emitting}. 
The method achieves state-of-the-art performance on word timing metrics, but its dependence on hybrid systems makes it less favored.
A purely E2E system was proposed where alignments from a hybrid system are not needed~\cite{chen21j_interspeech}.
But it requires training a frame-level classifier with CTC loss, manually expanding CTC peaks to word timings, and training another classifier using the word timings with cross-entropy (CE) loss.
Continuous integrate-and-fire (CIF) is a soft and monotonic alignment well suited for estimating word timings in E2E non-autoregressive ASR systems.
The CIF-based system obtains good acoustic boundaries by accumulating weights of frames, until exceeding a threshold value, and firing tokens.
With several steps of post-processing, its performance is comparable to HMM-based system and only need to be verified in languages other than Chinese~\cite{shi2023achieving}.


To the best of our knowledge, we are the first to adopt label priors to eliminate the CTC peaky behavior~\cite{zeyer2021does} for pure E2E word timing estimation and the proposed system surpasses the performance of HMM-based system on Chinese language.
Our contributions in this paper are:
\begin{enumerate} 
\item We adopt label priors to eliminate the peaky behavior of CTC and train frame-level classifiers to predict word timings, which can be easily extended to other languages.
\item We introduce Mel features as low-level local information and combine with Transformer encoder outputs as high-level utterance-wise information, which can be seen as feature fusion and is found to be beneficial for word timing estimation.
\item We further improve word timing accuracy by a frame-level knowledge distillation method to shift word boundaries adaptively.
\end{enumerate}
E2E speech recognition and word timing estimation systems are introduced in Section~2.
In Section~3, methods for optimizing the frame-level classifier used in this paper are presented.
Section~4 is the experimental setup.
Results and analyses are presented in Section~5, and a conclusion is given in Section~6.


\section{System Overview}
\label{sec:system_overview}

\begin{figure}[htb]
  \centering
  \includegraphics[width=0.8\linewidth]{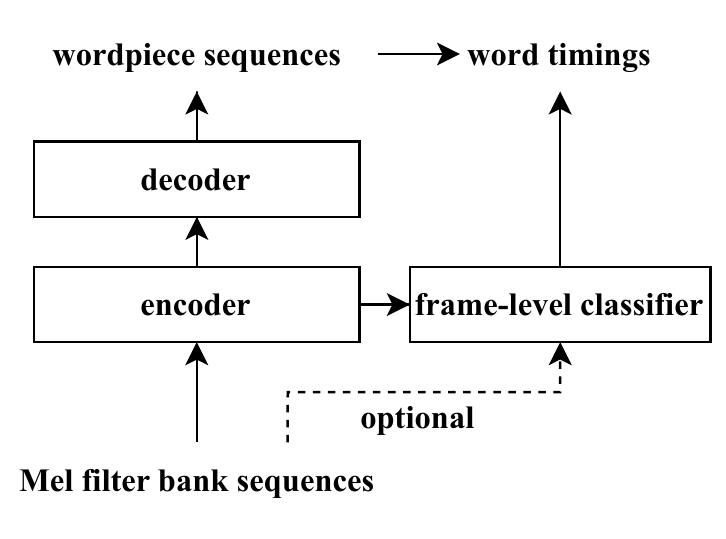}
  \vspace{-15pt}
  \caption{system workflow}
  \label{fig:e2e}
\end{figure}

The E2E system used in this paper is illustrated in Figure~\ref{fig:e2e}.
The ASR system employs LAS~\cite{chan2016listen} which is an attention-based encoder-decoder architecture.
The word timing system employs a frame-level classifier following the encoder and the Mel feature is introduced optionally.
The Mel filter bank sequence is given as $\mathbf{x}=(x_1,...,x_T)$ and the wordpiece sequence is given as $\mathbf{y}=(y_1,...,y_U)$, where $y_u$ is a wordpiece from a vocabulary of the size $V$.
The output of the decoder is denoted as $P(y_u|\mathbf{x},y_{<u})$ at the step $u$.
The output of the encoder is denoted as $\mathbf{h}=(h_1,...,h_T)$ and the encoder output with Mel feature denoted as $\mathbf{h}'=(h_1;x_1,...,h_T;x_T)$ (represented by +Mel).
The output of the frame-level classifier is denoted as $P(v,t|\mathbf{x})$ of the wordpiece $v$ at the time $t$.
The training stage consists of two steps.
First, the encoder-decoder is trained to maximize $P(y_u|\mathbf{x}, y_{<u})$.
Second, the parameters in the LAS are fixed and the frame-level classifier is trained using different methods which are described in the next section.
In the decoding stage, the encoder-decoder produces ASR results $\mathbf{y}'$ and the word timings come from the forced alignment between $\mathbf{y}'$ and $P(v,t|\mathbf{x})$.
\vspace{-5pt}
\vspace{-15pt}
\section{Methodology}

The frame-level classifier trained with CTC loss has peaky behavior where the peak roughly falls within the ground truth range of a token which can be character, wordpiece, or word.
The exact start and end times of a token can be obtained by estimating the left and right boundaries based on CTC peak position.
In this section, word timing estimation including peak position estimation and boundary estimation will be presented separately.
In addition, we also propose two methods to further polish word timing accuracy.

\subsection{Peak position estimation}
This subsection presents two methods combined to estimate the CTC peak position.

\begin{figure}[htb]
  \centering
  \includegraphics[width=\linewidth]{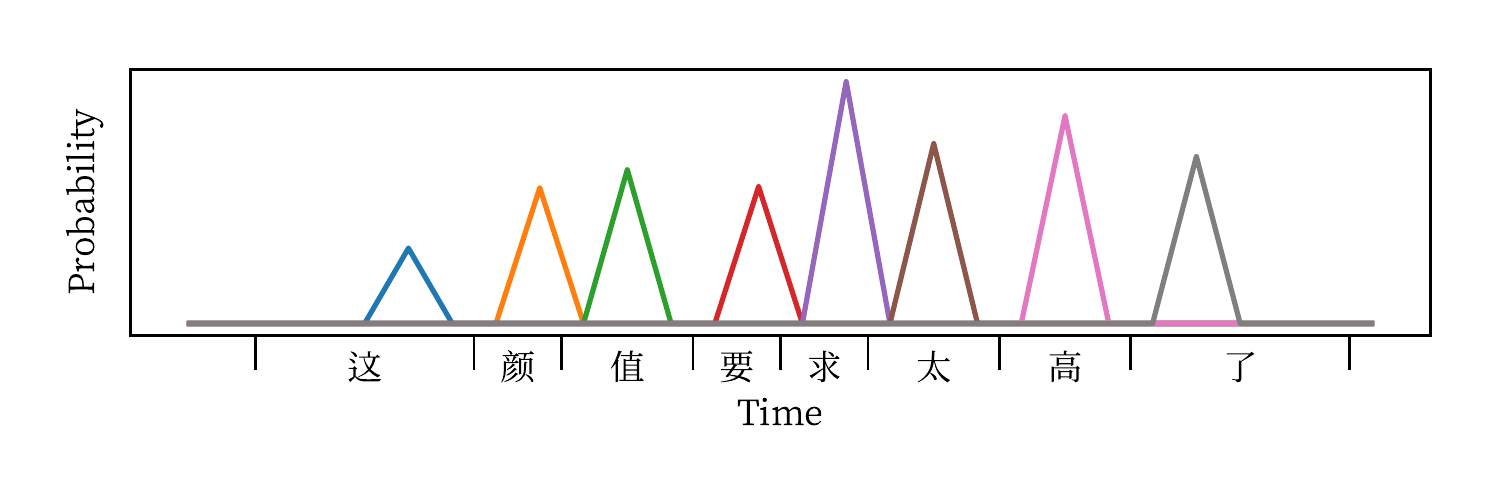}%
  \vspace{-15pt}
  \caption{CTC peaks}
  \label{fig:ctcpeak}
\end{figure}

CTC loss is proposed for training frame-level classifiers in E2E systems~\cite{graves2006connectionist}.
In order to align arbitrary frame-level and token-level sequences, the CTC model introduces a blank token (represented by $\phi$).
The CTC loss is computed by Equation~\ref{eq:ctc}, where $P(\pi_t|\mathbf{x})$ is the posterior probability of the token $\pi_t$ from the frame-level classifier at the time $t$, and the function $\mathcal{B}$ removes consecutive equivalent tokens as well as blank tokens so that $\mathcal{B}(\pmb{\pi})=\mathbf{y}$.
There are many valid paths $\pmb{\pi}$ that satisfy $\mathcal{B}(\pmb{\pi})=\mathbf{y}$, so the forward-backward algorithm is used to optimize the calculation of the CTC valid paths. 
Since the E2E system does not rely on the frame-level classifier to output ASR results, it must use forced alignment to obtain word timings.
Forced alignment, computed by Equation~\ref{eq:forcealignment}, aims to find the most likely valid path $\pmb{\pi}$.
The start and end time of each token $y_u$ is obtained through the path $\pmb{\pi}$.
Many experiments have shown that frame-level classifiers trained with CTC loss always predict blank tokens, known as the peaky behavior~\cite{liu2018connectionist, miao2015eesen, sak2015learning}.
As shown in Figure~\ref{fig:ctcpeak}, forced alignment of non-blank tokens last for few frames and peaks appear near the ground truths.
To summarize, the peaks can be obtained as anchor points for the position estimation by forced alignment from the CTC model.
\begin{equation}
  \mathcal{L}_\text{CTC} = -\log P(\textbf{y}|\textbf{x})=-\log \sum_{\pmb{\pi}\in\mathcal{B}^{-1}(y)}\prod_t P(\pi_t|\mathbf{x})
  \label{eq:ctc}
\end{equation}
\begin{equation}
  \pmb{\pi}=\operatorname*{argmax}_{\mathcal{B}(\pmb{\pi})=\mathbf{y}} P(\pmb{\pi}|\mathbf{x})=\operatorname*{argmax}_{\mathcal{B}(\pmb{\pi})=\mathbf{y}} \prod_t P(\pi_t|\mathbf{x})
  \label{eq:forcealignment}
\end{equation}
\vspace{-5pt}
\vspace{-15pt}

\subsection{Boundary estimation}
This subsection introduces two methods of peak expansion to estimate boundaries.

A previous method to estimate boundaries is based on neighboring CTC peaks, and teaches the frame-level classifier with a new posterior probability using cross-entropy~\cite{chen21j_interspeech}.
For convenience, this method is abbreviated to CETC.
The position of CTC peak of the token $y_u$ is denoted as $t_\text{peak}^u$.
The start and end times are computed by Equation~\ref{eq:cetc}.
The probability of the token $y_u$ at the time $t$ is $P(y_u,t|\mathbf{x})$ from the classifier trained with $\mathcal{L}_\text{CETC}=\text{Cross-Entropy}(P,P_\text{guided})$, where $P_\text{guided}$ is computed by Equation~\ref{eq:guided}.
The hyperparameters $\alpha_\text{left}$ and $\alpha_\text{right}$ control the left and right boundaries, and $\beta$ controls that the further away from the CTC peak, the lower the probability.
According to the CETC paper, the $(\alpha_\text{left},\alpha_\text{right},\beta)$ are set to $(0.2,0.7,0.5)$.
The CETC method estimating boundaries by neighboring peaks is used to estimate word timings.
\begin{equation}
\begin{split}
  t_\text{start}^{u} &= t_\text{peak}^{u}-\alpha_\text{left} \cdot (t_\text{peak}^{u}-t_\text{peak}^{u-1}) \\
  t_\text{end}^{u} &= t_\text{peak}^{u}+\alpha_\text{right} \cdot (t_\text{peak}^{u+1}-t_\text{peak}^{u})
  \label{eq:cetc}
\end{split}
\end{equation}
\begin{equation}
\begin{split}
  P_\text{guided}(y_u,t|\mathbf{x})&=\left\{
\begin{array}{ccl}
(\frac{t-{t}_\text{peak}^{u}}{{t}_\text{peak}^{u}-{t}_\text{start}^{u}})^\beta , & {t_\text{start}^{u} \leq t < t_\text{spike}^{u}} \\
(\frac{{t}_\text{end}^{u}-t}{{t}_\text{end}^{u}-{t}_\text{peak}^{u}})^\beta , & {t_\text{spike}^{u} \leq t \leq t_\text{end}^{u}} 
\end{array} \right. \\
P_\text{guided}(\phi,t|\mathbf{x}) &= 1-\sum P(\neq\phi,t|\mathbf{x})
  \label{eq:guided}
\end{split}
\end{equation}

A CTC model without peaky behavior is well suited for boundary estimation.
A recent paper explains and elegantly solves the CTC peaky behavior by introducing the label prior, which is abbreviated to the non-peaky CTC (NPC) method here~\cite{npc}.
The label prior must be calculated using logits before softmax function, computed by Equation~\ref{eq:prior}.
The modified logits are computed by Equation~\ref{eq:npc} and the NPC paper suggests using different $\gamma$ in training and inference, which will be reported in the following section.
The CTC loss is calculated on the modified logits.
The NPC method extending a peak to a span by reducing the probability of blank token is employed to estimate the token boundaries.
\begin{equation}
\begin{split}
  O_\text{prior}(y_u|\mathbf{x}) = \frac{1}{T} \sum_{t=1}^{T} O(y_u,t|\mathbf{x})
\end{split}
 \label{eq:prior}
\end{equation}
\begin{equation}
\begin{split}
  O'(y_u,t|\mathbf{x}) = O(y_u,t|\mathbf{x}) - \gamma \cdot O_\text{prior}(y_u|\mathbf{x})
\end{split}
 \label{eq:npc}
\end{equation}

\subsection{Polishing word timing accuracy}

In our experiments, we found that the NPC method tends to estimate word boundaries earlier in time, which leaves room for improvement in word timing accuracy.
Having the manually labelled development set, the best shifting for predicted boundaries can be grid searched and applied in post-processing.
On the other hand, to shift word timings adaptively, the CTC peaks are shifted using a frame-by-frame knowledge distillation method called Peak First Regularization (PFR)~\cite{tian2022peak}.
The PFR loss is computed by Equation~\ref{eq:pfr}, where $\mu=-1$ delays CTC peaks, $\mu=1$ advances CTC peaks and $P_\text{PFR}$ is the processed classifier probability.
A temperature coefficient $\tau$ in Equation~\ref{eq:tau} smooths the output of the frame-level classifier $O(v,t|\mathbf{x})$ of the token $v$ at the time $t$.
According to the independence assumption of CTC, the PFR loss must be employed before freezing the LAS and the loss is computed by Equation~\ref{eq:loss}.
It is necessary to prevent the gradient from passing through the teacher frame $P_\text{PFR}(v,t+\mu|\mathbf{x})$.
The default $(\mu,\tau,\lambda_\text{CE})$ is $(-1,10.0,0.95)$ when using the PFR loss.
\begin{equation}
  \mathcal{L}_\text{PFR} = \sum_t \sum_v P_\text{PFR}(v,t+\mu|\mathbf{x})\log \frac{P_\text{PFR}(v,t+\mu|\mathbf{x})}{P_\text{PFR}(v,t|\mathbf{x})}
  \label{eq:pfr}
\end{equation}
\begin{equation}
  P_\text{PFR}(v,t|\mathbf{x})=\text{softmax}(O(v,t|\mathbf{x})/\tau)
  \label{eq:tau}
\end{equation}
\begin{equation}
  \mathcal{L} = \lambda_\text{CE} \cdot \mathcal{L}_\text{CE} + (1 - \lambda_\text{CE}) \cdot (\mathcal{L}_\text{CTC} + \lambda_\text{PFR} \cdot \mathcal{L}_\text{PFR})
  \label{eq:loss}
\end{equation}

\section{Experiments}

\subsection{Datasets}
We use the LibriSpeech 960h for base experiment and apply the word timing estimation method to the internal data in 7 languages including Chinese (zh), English (en), German (de), Russian (ru), Spanish (es), Indonesian (id) and Vietnamese (vi) where the total durations are 100k, 40k, 9k, 10k, 20k, 6k and 10k hours respectively.
The word timing test set for LibriSpeech is the forced alignment result from HMM-based hybrid system with script\footnote{https://github.com/kaldi-asr/kaldi/blob/master/egs/librispeech/s5/run.sh} and the internal test sets are annotated by the experts.
The E2E system employs 80-dimensional Mel-filter bank features as input and wordpiece sequences obtained with SentencePiece tokenizer~\cite{spm2} as output.

\subsection{Modeling}
In this paper, an encoder-decoder model outputs ASR results and a frame-level classifier of 2 dense layers with 2048 hidden units following the encoder outputs word timings.
The 80-dimensional Mel-filter bank is computed on a \SI{25}{\milli\second} window with a \SI{10}{\milli\second} shift.
In the encoder, two CNN kernels of width, height, input channel size, and output channel size $(3,3,1,64)$ and $(3,3,64,128)$ with $(2,2)$ stride are used to extract high-level feature.
Then the CNN feature projected to 512-dimension is fed into 18 transformer blocks.
The transformer blocks have a multi-head self-attention layer with 8 heads and a feed-forward layer adopting GLU activation function with 2048 hidden units.
The output of the last transformer block is projected from 512 to 2048.
The decoder has a multi-head attention layer with 2 heads, 4 LSTM layers with 1024 hidden units and an embedding layer with 512 hidden units.
The encoder output optionally combines 4 consecutive frames of the Mel features into 2368-dimensional features, which are fed to the frame-level classifier as acoustic features.
Overall the model has 168M parameters.

The E2E system is trained in Tensorflow~\cite{tensorflow} using the Lingvo~\cite{lingvo} toolkit and RETURNN~\cite{returnn} toolkit (for calculating the modified CTC loss) on 16 Tesla A100 GPUs for around 5 days.
The hybrid system is trained in Kaldi~\cite{kaldi}.

\subsection{Configuration}
Using Adam optimizer to train the LAS model, the highest learning rate is $7.5\mathrm{e}{-4}$ at the end of the warm-up at 4k steps.
The learning rate starts decaying at 60k steps decaying by half every 50k steps.
A model is trained 200k steps, and models for high-resource languages are trained up to 300k steps.
Audios of similar length are grouped into same batches, and each step of batch size multiplied by the longest audio in the batch is approximately 1224 seconds.
The training technique is from the LAS paper~\cite{chan2016listen}, using Specaugment~\cite{park19e_interspeech} where $(F,m_F,T,p,m_T)=(27,2,100,0.1,1)$.
The variational weight noise~\cite{vn} is turned on optionally where standard deviation is 0.075 and the uniform label smoothing~\cite{labelsmooth} is introduced with uncertainty 0.1.
In addition MWER~\cite{mwer} and CTC fine tuning with fixed learning rates $1\mathrm{e}{-6}$ are trained for 20k and 10k steps respectively.

\subsection{Measuring word timings}
\newcommand\ST{\text{Ave. ST$\Delta$}}
\newcommand\ED{\text{Ave. ED$\Delta$}}
\newcommand\WS[1]{\text{\%WS\textless{}#1ms}}
\newcommand\WE[1]{\text{\%WE\textless{}#1ms}}
The word timing metrics are adopted from~\cite{sainath2020emitting}, which are the average offset (ms) of the predicted word start to the ground truth (\ST) and the percentage of those word start offsets smaller than \SI{200}{\milli\second} (\WS{200}).
Similarly, the \ED~and \WE{200} for word end.
In our experiments, word timing metrics are only calculated on the matched token pairs according to the alignment of minimum edit distance.
For higher granularity we employ \WS{80} and \WE{80} in our experiments.

\section{Results}







\subsection{Analyzing the label prior}
Table~\ref{tab:prior} shows the word timing metrics with different label priors used in training and inference on LibriSpeech test-clean.
The frame-level classifier which is trained with CTC loss predicts mostly blanks and rarely the other tokens, known as peaky behavior.
The model with peaky behavior has bad word timing metrics even though inference with the label prior when $\gamma_\text{train}=0, \gamma_\text{inf} \neq 0$.
It shows that the word timing metrics have a significant improvement because of training with the label prior as a normalization term when $\gamma_\text{train}\neq0$.
In the following experiments, the default $(\gamma_\text{train},\gamma_\text{inf})$ are (0.25, 1.0) because the summation of the \WS{80} and the \WE{80} is the largest.

\begin{table}[htb]
\setlength\tabcolsep{3pt}
\centering
\caption{Effect of the label prior ($\gamma$) in training and inference on word timing metrics on LibriSpeech}
\label{tab:prior}
\begin{tabular}{@{}cccccc@{}}
\toprule
$\gamma_\text{train}$ & $\gamma_\text{inf}$ & \ST    & \ED    & \WS{80} & \WE{80} \\ \midrule
0.0                  & 0.0                & 117.51 & 158.06 & 60.57   & 44.59   \\
0.0                  & 1.0                & 103.21 & 148.25 & 69.05   & 50.31   \\
0.25                 & 0.0                & 41.36  & 52.92  & 96.71   & 90.19   \\
0.25                 & 1.0                & 40.28  & 50.54  & \textbf{96.75}   & \textbf{91.18}   \\
0.5                  & 0.0                & 42.4   & 50.35  & 95.48   & 91.35   \\
0.5                  & 1.0                & 42.54  & 49.18  & 95.14   & 91.87   \\
0.75                 & 0.0                & 47.36  & 51.4   & 93.01   & 90.51   \\
0.75                 & 1.0                & 48.11  & 50.77  & 92.4    & 90.63   \\
1.0                  & 0.0                & 49.89  & 52.61  & 91.66   & 89.57   \\
1.0                  & 1.0                & 50.49  & 51.55  & 91.0    & 89.64   \\ \bottomrule
\end{tabular}
\end{table}
\vspace{-10pt}
\subsection{Improving frame-level classifiers in 7 languages}
Table~\ref{tab:business} shows the performance of 4 frame-level classifiers in Chinese and compares the proposed method NPC+Mel to the previous work CETC in 6 other languages. 
HMM denotes the frame-level classifier from HMM-based hybrid system, and Offset denotes the offset in \SI{}{\milli\second} applied to all timestamps.
It shows that NPC outperforms HMM in Chinese when \SI{40}{\milli\second} offset is applied.
Further, the introduction of Mel feature is to provide accurate local information to reduce the amount of offset needed.
Nevertheless it doesn't meet the purpose, it improves performance.
Hence, we suspect the cause of the word timing shift is not from feature extraction but the use of CTC loss.
NPC+Mel obtains an average of 86.79\%/78.99\% for \WS{80} and \WE{80} in all 7 languages, greatly outperforms CETC's 82.14\%/70.94\% and slightly surpasses NPC's 86.53\%/78.10\%\footnote{Due to space limitation, the details of NPC results for 6 languages are not listed in Table~\ref{tab:business}}.

\begin{table}[htb]
\setlength\tabcolsep{2pt}
\centering
\caption{Word timing metrics in 7 languages in internal datasets}
\label{tab:business}
\resizebox{1\columnwidth}{!}{
\begin{tabular}{@{}ccccccc@{}}
\toprule
Method  & Lang & \ST    & \ED    & \WS{80} & \WE{80} & Offset \\ \midrule
HMM     & zh  & 41.47  & 47.05  & 93.0    & 90.22   & 0     \\
CETC    & zh  & 39.62  & 44.53  & 91.86   & 89.72   & 0     \\
NPC     & zh  & 32.69  & 35.39  & 95.5    & 93.07   & 40    \\
NPC+Mel & zh  & 31.93  & 33.78  & \textbf{95.68}   & \textbf{94.18}   & 40    \\ \midrule
CETC    & en    & 55.66  & 105.43 & 78.72   & 58.33   & 0     \\
NPC+Mel & en    & 44.05  & 70.35  & \textbf{88.64}   & \textbf{77.91}   & 60    \\ \midrule
CETC    & ru    & 53.6   & 92.63  & 82.18   & 61.62   & 0     \\
NPC+Mel & ru    & 48.49  & 80.49  & \textbf{85.2}    & \textbf{69.68}   & 60    \\ \midrule
CETC    & es    & 180.58 & 189.14 & 59.72   & 55.5    & 0     \\
NPC+Mel & es   & 173.7  & 181.4  & \textbf{62.53}   & \textbf{60.62}   & 70    \\ \midrule
CETC    & de    & 44.78  & 63.0   & 86.01   & 75.94   & 0     \\
NPC+Mel & de    & 33.62  & 52.64  & \textbf{93.61}   & \textbf{83.7}    & 60    \\ \midrule
CETC    & id    & 49.3   & 80.85  & 85.53   & 69.98   & 0     \\
NPC+Mel & id    & 46.23  & 80.62  & \textbf{88.11}   & \textbf{76.28}   & 60    \\ \midrule
CETC    & vi    & 38.48  & 47.92  & 90.99   & 85.49   & 0     \\
NPC+Mel & vi    & 30.09  & 39.62  & \textbf{94.85}   & \textbf{90.36}   & 40    \\ \bottomrule
\end{tabular}
}
\end{table}
\vspace{-10pt}
\subsection{Analyzing CTC peak positions}
\begin{figure}[htb]
  \centering
  \includegraphics[width=\linewidth]{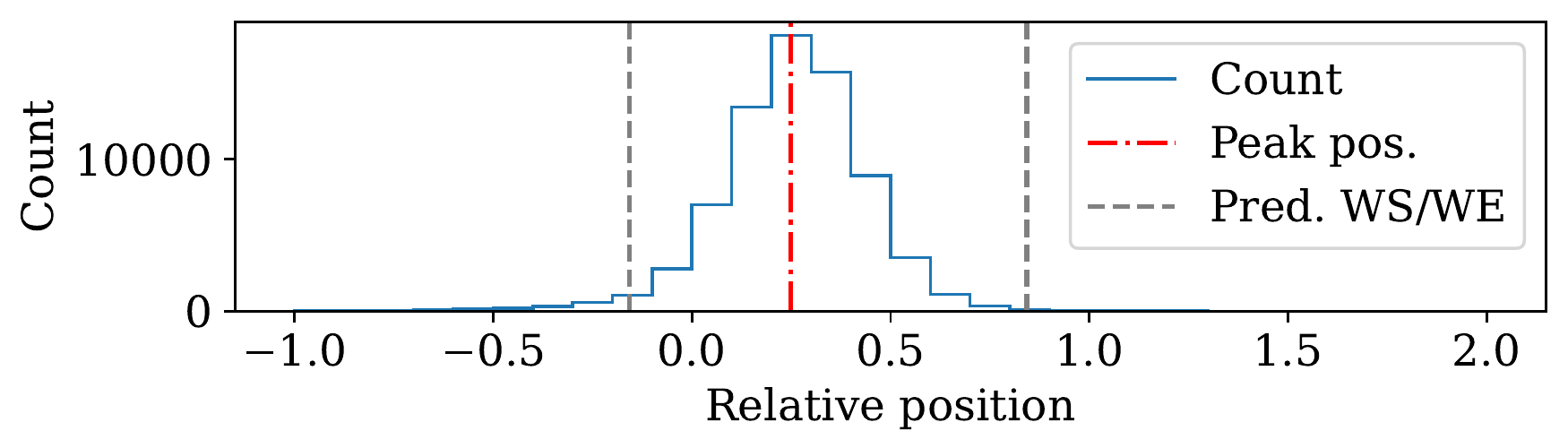}
  \vspace{-15pt}
  \caption{Distribution of the relative position of CTC peaks}
  \label{fig:peaks}
\end{figure}

The introduction of CTC loss brings the peaky behavior, so we first analyze the CTC peak positions.
As shown in Figure~\ref{fig:peaks}, the number of CTC peaks is measured as y-axis and the relative position of the peak time to the ground truth word timings is measured as x-axis.
$x=0$ means that CTC peaks are emitted on the word start time and $x=-1$ means that CTC peaks appear a whole word duration early before the start time.
The red dotted dashed line in the figure shows the average relative position of the peak to the ground truth is 0.249.
The grey dashed lines indicate the average word start and end times of the NPC+Mel prediction, and the CTC peaks are located on average at 0.406 of the relative position of the prediction.
The reason that applying positive offset to the predicted timestamps increases accuracy is because the predicted word start and end times are usually earlier than they should be, as shown in Figure~\ref{fig:peaks}, the grey dashed lines do not lie exactly on $x=0$ and $x=1$.
Table~\ref{tab:duration} shows that the average word duration of the prediction is very close to the ground truth.
In conclusion, the NPC+Mel method accurately predicts the word duration, and by delaying the emission time of CTC peaks we might reduce the amount of offset needed for increasing accuracy.

\vspace{-15pt}
\begin{table}[htb]
\setlength\tabcolsep{3pt}
\centering
\caption{Average word duration of the ground truth and prediction (\SI{}{\milli\second}) in internal datasets}
\label{tab:duration}
\resizebox{1\columnwidth}{!}{
\begin{tabular}{lccccccc}
\toprule
             & zh & en & ru & es & de & id & vi \\ \midrule
Ground truth & 163.3 & 287.1 & 376.6 & 287.7 & 289.5 & 368.0 & 201.0 \\
Prediction      & 162.8 & 257.2 & 331.6 & 288.7 & 279.3 & 337.5 & 195.8 \\ \bottomrule
\end{tabular}
}
\end{table}
\vspace{-10pt}
\subsection{Delaying CTC peaks and timestamps}
We conduct the experiment in LibriSpeech.
The CTC peak is delayed by introducing PFR loss during training.
Table~\ref{tab:pfr} shows the performance with different $\lambda_\text{PFR}$.
The E2E system is trained from scratch with CE, CTC and PFR loss.
Because of the independence assumption in CTC loss, PFR loss cannot change the position of CTC peaks in the case of fixed encoder parameters.
Without fixing the encoder parameters, the ASR result is affected and we observe WER increases greatly as $\lambda_\text{PFR}$ surpasses $2.5$.
The result shows that setting $\lambda_\text{PFR}=1.5$ not only reduces offset to \SI{0}{\milli\second} but also achieves best word timing metrics.

\begin{table}[htb]
\setlength\tabcolsep{3pt}
\centering
\caption{Effect of $\lambda_\text{PFR}$ on the offset needed to achieve the highest accuracy on LibriSpeech}
\label{tab:pfr}
\resizebox{1\columnwidth}{!}{
\begin{tabular}{lccccrr}
\toprule
$\lambda_\text{PFR}$ & \ST   & \ED   & \WS{80} & \WE{80} & Offset & WER  \\ \midrule
0.0       & 40.28 & 50.54 & 96.75   & 91.18   & 40    & 5.3  \\
0.5       & 35.06 & 42.04 & 97.35   & 93.84   & 20    & 4.1  \\
1.0       & 34.73 & 40.85 & 98.0    & 95.18   & 10    & 4.7 \\
1.5       & 38.69 & 42.9  & \textbf{97.99}   & \textbf{96.04}   & 0     & 5.0 \\
2.0       & 32.57 & 35.96 & 97.89   & 94.82   & -10   & 4.2 \\
2.5       & 40.36 & 41.81 & 97.84   & 95.59   & -20   & 4.7 \\
3.0       & 51.62 & 55.69 & 97.67   & 93.96   & -30   & 6.3 \\ \bottomrule
\end{tabular}
}
\end{table}
\vspace{-10pt}
\section{Conclusions}
In this paper, we improve frame-level classifiers with non-peaky CTC loss in a purely E2E system.
The proposed method outperforms the HMM-based hybrid system in Chinese and the CETC method in 7 languages.
We observe the timestamp accuracy can be further improved, which requires manually labelled development set for gridsearching the offset of timestamp in post-processing.
Also we employ PFR method to adaptively shift CTC peaks in training, improving word timing performance.
In the future, the proposed approach can be applied to a multilingual E2E ASR system for word timings.




\bibliographystyle{IEEEtran}
\bibliography{main}

\end{document}